**Epitaxial Magnon Junctions with $Y_3Fe_5O_{12}$/$Gd_3Ga_5O_{12}$/Pt-nanoparticles/$Y_3Fe_5O_{12}$**

*Jianing Lin, Haoyu Zhuang, Tianyi Zhang, Jiyang Ou, Jihao Xia, Hongyu Wang, Jiang Niu, Qinglong Yu, Caihua Wan*, Guoqiang Yu, Richeng Yu, and Xiufeng Han**

Jianing Lin, Haoyu Zhuang, Tianyi Zhang, Jiyang Ou, Jihao Xia, Hongyu Wang, Jiang Niu, Qinglong Yu, Caihua Wan, Guoqiang Yu, Richeng Yu, and Xiufeng Han
Beijing National Laboratory for Condensed Matter Physics, Institute of Physics, Chinese Academy of Sciences, Beijing 100190, China
E-mail: xfhan@iphy.ac.cn; wancaihua@iphy.ac.cn

Jianing Lin, Tianyi Zhang, Jiyang Ou, Jihao Xia, Hongyu Wang, Jiang Niu, Qinglong Yu, Guoqiang Yu, and Xiufeng Han
Center of Materials Science and Optoelectronics Engineering, University of Chinese Academy of Sciences, Beijing 100049, China

Haoyu Zhuang, and Richeng Yu
School of Physical Sciences, University of Chinese Academy of Sciences, Beijing 100049, China

Jiyang Ou, Caihua Wan, Richeng Yu, and Xiufeng Han
Songshan Lake Materials Laboratory, Dongguan, Guangdong 523808, China

**Abstract**

The excitation, propagation, manipulation, and detection of magnons lie at the core of magnonics. Among various device architectures, high-quality magnon junctions with a ferromagnetic insulator (FI)/antiferromagnetic insulator (AFI)/ferromagnetic insulator (FI) configuration are predicted to exhibit a near-100% magnon junction ratio (MJR), making them promising candidates for high-performance non-volatile memory applications. Yttrium iron garnet ($Y_3Fe_5O_{12}$, YIG) is widely employed as the FI material in such structures for its exceptionally low magnetic damping. However, in previously reported magnon junctions, the epitaxial relationship between the two YIG layers is often disrupted by lattice-mismatched AFI spacers. Although this leads to distinct coercivities, it simultaneously degrades the crystalline quality of the top YIG layer, thereby limiting the magnontransmission efficiency. Here, an epitaxial magnon junction is demonstrated incorporating a hybrid $Gd_3Ga_5O_{12}$ (GGG)/Pt-nanoparticles spacer that maintains coherent epitaxy while enabling an antiparallel magnetic configuration. Compared with nickel oxide-based magnon junctions, the coercivity-tuned top YIG layer in our structure exhibits superior crystalline quality and much lower magnetic damping. A clear magnon junction effect is observed with a 43% MJR achieved. The interlayer coupling between the two ferrimagnetic insulators is found to enhance the MJR. These results offer valuable insights for the development of advanced magnon-based spintronic devices.

## 1. Introduction

Spin waves are collective excitations in magnetically ordered systems, and magnons are the quantized quasiparticles of these spin waves.[1–3] Unlike spin-polarized electron currents, magnons can propagate through magnetic insulators without involving charge transport, thereby eliminating Joule heating during information transmission. Consequently, employing magnons as information carriers holds great promise for realizing ultra-low-power spintronic or magnonic devices.[4] Driven by this potential, the field of magnonics has emerged in recent years, focusing on the excitation, propagation, manipulation, and detection of magnons. A variety of magnonic devices have been proposed and demonstrated, including magnon valves,[5–7] magnon junctions,[8–10] magnon transistors,[11–13] magnonic directional couplers,[14] magnon crystal,[15] and so on, which together provides a rich platform for developing magnon-based information technologies.

Magnon valves and magnon junctions are significant devices in the field of magnonics. Wu et al. first proposed in 2018 a $Y_3Fe_5O_{12}$(YIG)/Au/$Y_3Fe_5O_{12}$(YIG) magnon valve and observed a clear magnon valve effect — the modulation of magnon current by varying the relative orientation of the magnetic moments in the top and bottom layer.[5] Vilela et al. subsequently realized a $Tm_3Fe_5O_{12}$/Au/$Tm_3Fe_5O_{12}$ magnon valve with perpendicular magnetic anisotropy.[6] Truong et al. further demonstrated a magnon valve incorporating a Pt spacer.[7] More important, Guo et al. first proposed in 2018 all-oxide and insulating magnon junctions employing an antiferromagnetic insulator as the spacer, in which spin current is only delivered by pure magnons instead of any diffusive spin current. Their works experimentally revealed both a distinct magnon junction effect and a nonlocal spin Hall magnetoresistance effect.[8–10] Further, Zhang et al. in 2023 predicted that near 100% magnon junction ratio (MJR) in magnon junctions with a full quantum model.[16] However, in most of the previously reported magnon valves and magnon junctions, the antiparallel (AP) magnetic configuration between the top and bottom YIG layers was realized by exploiting differences in coercivity arising from variations in crystalline quality induced by the lattice-mismatched spacer.[17–19] In such cases, the degraded crystalline quality severely hinders efficient magnon transport by increasing magnetic damping. [20-22]

Gadolinium gallium garnet ($Gd_3Ga_5O_{12}$, GGG) is a paramagnetic material that shares the same crystal structure with YIG and exhibits an extremely small lattice mismatch of only 0.057%.[23–24] Zhu et al. investigated magnon transport in YIG/GGG/Pt heterostructures and discovered that paramagnetic magnons in GGG can also mediate angular momentum transfer at room temperature and small fields.[25] Inspired by this finding, we proposed that paramagnetic GGG can serve as an ideal spacer in magnon junctions, decoupling the magnetic interaction between adjacent YIG layers while maintaining the overall epitaxial relationship. Based on this concept, we have innovatively designed a YIG/GGG/Pt-nanoparticles/YIG magnon junction. During high-temperature annealing, Pt films deposited on dielectric materials undergo dewetting, transforming into Pt nanoparticles. [26–27] By precisely controlling the nominal thickness of deposited Pt, the coercivity of the upper YIG layer can be finely tuned, enabling the realization of an AP magnetic configuration of the two YIG layers. The gaps between Pt nanoparticles further facilitate the continuation of epitaxial growth, resulting in a high-quality magnon junction. Ferromagnetic resonance (FMR) measurements confirm that YIG grown on a GGG/Pt-nanoparticles spacer exhibits magnetic damping comparable to that of epitaxial YIG directly grown on GGG substrate, and remarkably, the damping is reduced by a factor of 22.5 compared with samples using nickel oxide (NiO) as the spacer. Finally, we successfully observed a distinct magnon junction effect in the magnon junction with the GGG/Pt-nanoparticles spacer, achieving a MJR of 43%. By comparing the temperature dependence of the MJR with that of the coercivity of the top YIG layer, a clear correlation is observed, indicating that interlayer coupling between the two ferrimagnetic insulators contributes to an increase in the MJR. Hopefully, this YIG/GGG/Pt-nanoparticles/YIG design will shed light on the realization of low-loss magnon junctions with high MJR, thereby enabling their application in magnonics and advanced information technologies.

## 2. Results and Discussion

We fabricated YIG(40)/GGG($t_{\mathrm{GGG}}$)/Pt($t_{\mathrm{Pt}}$)/YIG(30) heterostructures on (111)-oriented GGG substrates (sub.) using magnetron sputtering, where the numbers in parentheses after denote the nominal thicknesses of deposited films in nanometers before annealing. Here, $t_{\mathrm{GGG}}$ and $t_{\mathrm{Pt}}$ represent the nominal thicknesses of the deposited GGG and Pt layers before annealing, respectively. The as-deposited films were subsequently annealed in air at 800 °C for 80 minutes. **Figure 1**(a) presents a schematic illustration of the multilayer structure, where T-YIG and B-YIG denote the top and bottom YIG layers, respectively. After high-temperature annealing, the sputtered Pt layer undergoes dewetting and transforms into discontinuous islands in the form of nanoparticles.[26–27] Electrical transport measurements confirmed that the film is insulating after annealing (as shown in **Figure S1**). Figure 1(b) presents a cross-sectional high-angle annular dark-field (HAADF) image of the as-deposited YIG(40)/GGG(5)/Pt(1)/YIG(30) heterostructure, showing a continuous Pt spacer. After annealing, as shown in Figure 1(c), the Pt spacer transforms into embedded nanoparticles within the GGG matrix. The interfacial structures from the T-YIG layer to the GGG/Pt-nanoparticles spacer and from the GGG/Pt-nanoparticles spacer to the B-YIG layer are displayed in Figures 1(d) and 1(e), demonstrating that the presence of Pt nanoparticles does not disrupt the epitaxial relationship among B-YIG, GGG, and T-YIG. In addition, **Figure S2**(a) confirms the epitaxial growth from the GGG substrate to the B-YIG layer, indicating that the overall epitaxial coherence of the full stack is well maintained. Figure S2(b) shows the electron energy loss spectroscopy (EELS) elemental maps, revealing partial diffusion of Y and Fe into the intermediate GGG layer, whereas no noticeable diffusion of Gd or Ga into either the B-YIG or T-YIG layers is detected. A sharp interface between the GGG substrate and the B-YIG layer is preserved.

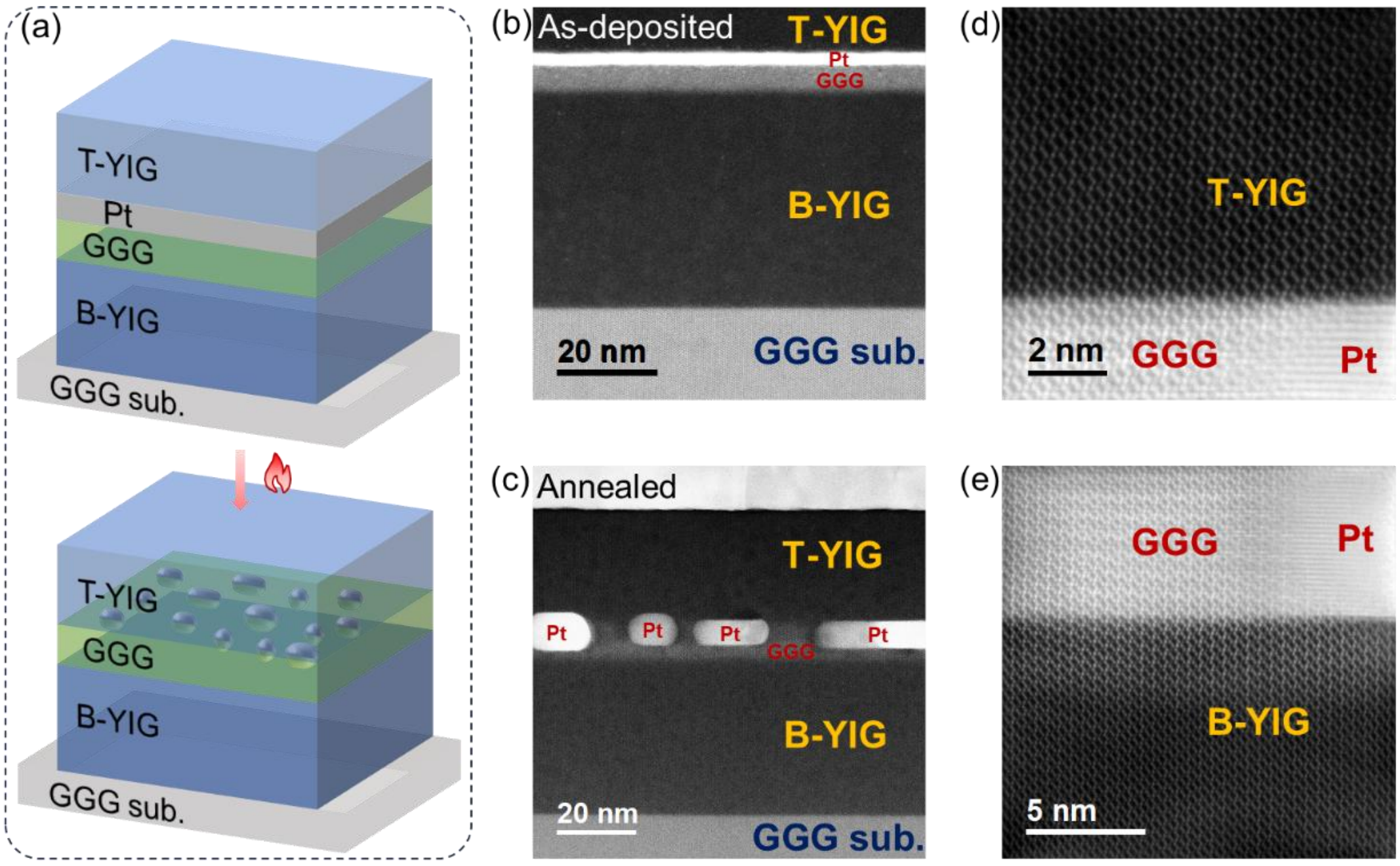

**Figure 1.** (a) Schematic illustration of the annealing process for the deposited GGG//YIG(40)/GGG($t_{GGG}$)/Pt($t_{Pt}$)/YIG(30) heterostructure. (b) Cross-sectional HAADF image of the as-deposited GGG//YIG(40)/GGG(5)/Pt(1)/YIG(30) film, showing a continuous Pt layer. (c) Cross-sectional HAADF image of the annealed film, demonstrating that the continuous Pt layer transforms into Pt nanoparticles after annealing. (d) Zoomed-in HAADF image of the interface between the T-YIG layer and the GGG/Pt-nanoparticles spacer in the annealed sample, where the T-YIG maintains high crystallinity and preserves epitaxial alignment with the underlying GGG layer. (e) Zoomed-in HAADF image of the interface between the GGG/Pt-nanoparticles spacer and the B-YIG layer, showing that the GGG spacer also retains high crystallinity and maintains epitaxy with the adjacent B-YIG.

Magnetic characterization of the films was performed using a vibrating sample magnetometer (VSM). With the GGG spacer thickness fixed at $t_{GGG}$ = 5 nm, the magnetic hysteresis loops (black solid lines) and the corresponding differential curves (red dashed lines) for $t_{Pt}$ = 0.5, 0.7, 0.9, and 1 nm are shown in **Figure 2**(a). The coercive fields of the two YIG layers become clearly distinguishable when $t_{Pt}$ reaches 1 nm. Similar behavior is observed for $t_{GGG}$ = 6 and 7 nm, as presented in Figure 2(b), where the coercivity difference between the two YIG layers also allows the formation of distinct magnetic states, parallel (P) or antiparallel (AP) ones. For comparison, **Figure S3** shows the hysteresis loops of GGG//YIG(40)/GGG($t_{GGG}$)/YIG(30) films without the Pt-nanoparticles spacer. In this case, even when the thickness of the inserted GGG layer is increased up to 16 nm which far exceeds the magnon transport length in GGG,[25] the coercive fields of the two YIG layers remain nearly identical, making it impossible to achieve the AP magnetic configuration by applying external magnetic field.

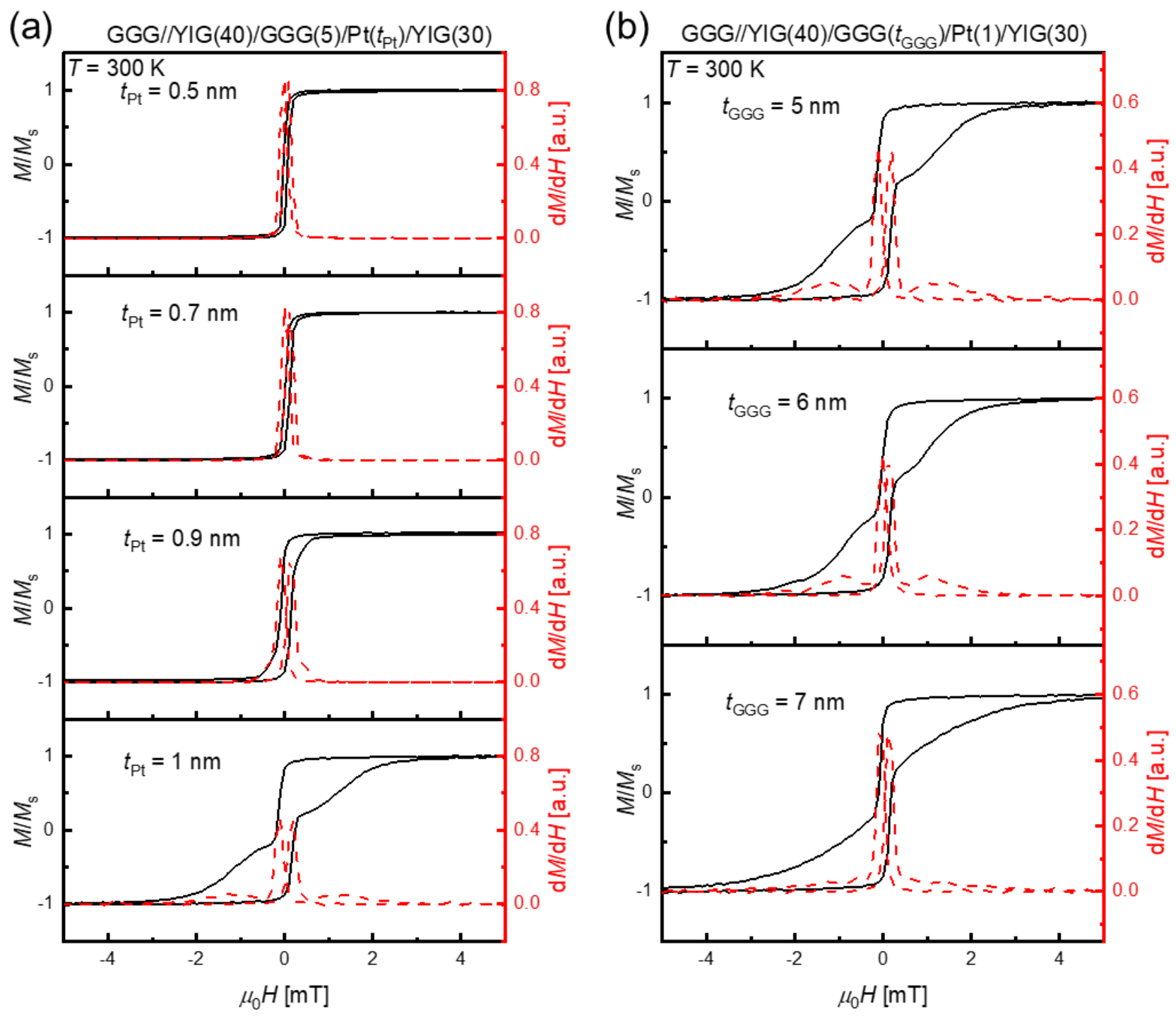

**Figure 2**. (a) Magnetic hysteresis loops of the annealed YIG(40)/GGG(5)/Pt($t_{Pt}$)/YIG(30) films with fixed $t_{GGG}$ = 5 nm and varying $t_{Pt}$ = 0.5, 0.7, 0.9, and 1 nm at 300 K. When $t_{Pt}$ = 1 nm, the two YIG layers exhibit distinguishable coercive fields, enabling the formation of an AP magnetic configuration. (b) Magnetic hysteresis loops for films with fixed $t_{Pt}$ = 1 nm and varying $t_{GGG}$ = 5, 6, and 7 nm at 300 K. AP states are observed in all three cases, confirming the effectiveness of Pt nanoparticles in tuning the coercivity of the T-YIG layers.

To evaluate the crystalline quality and magnetic damping of the coercivity-tuned T-YIG layer, three types of samples — GGG(5)/Pt(1)/YIG(30), YIG(30), and NiO(8)/YIG(30) — were deposited on (111)-oriented GGG substrates and then annealed under the identical condition. FMR measurements were performed to characterize their magnetic dynamic properties. Using field modulation lock-in technique, the detected spectrum is proportional to the derivative of the microwave absorption power with respect to the static magnetic field.[28–30] **Figure 3**(a) shows the FMR spectra of the three samples measured at a microwave frequency of 10 GHz. The resonance field ($\mu_0 H_r$) and linewidth ($\mu_0 \Delta H$) were extracted by fitting the data using the derivative of an asymmetric Lorentzian function, where $\mu_0 \Delta H$ is defined as the half-width at half-maximum (HWHM) of the FMR power absorption peak and $\mu_0$ represents the vacuum permeability.[29] The obtained $\mu_0 \Delta H$ are 6.50 ± 0.03 mT for annealed GGG(5)/Pt(1)/YIG(30) sample, 1.04 ± 0.02 mT for annealed YIG(30) sample, and 19.9 ± 0.1 mT for annealed NiO(8)/YIG(30) sample. The frequency dependence of the resonance field was further analyzed using the Kittel formula:[28–30]

$$f = |\gamma| \mu_0 \sqrt{H_r (H_r + M_s - H_a)} \quad (1)$$

where $f$ denotes the microwave frequency, $|\gamma|$ is the absolute value of the gyromagnetic ratio, $M_s$ is the saturation magnetization, and $H_a$ represents the effective anisotropy field. The fitting results are shown in Figure 3(b). The extracted $|\gamma|$ values for the three samples are 27.81 ± 0.04 MHz mT$^{-1}$ for annealed GGG(5)/Pt(1)/YIG(30) sample, 28.13 ± 0.02 MHz mT$^{-1}$ for annealed YIG(30) sample, and 27.7 ± 0.2 MHz mT$^{-1}$ for annealed NiO(8)/YIG(30) sample, which are in excellent agreement with previously reported values.[28][30] Using the obtained $|\gamma|$ values, the Gilbert damping constant ($\alpha$) of each film can be determined by fitting the frequency dependence of the FMR linewidth according to the following relation:[28–30]

$$\mu_0 \Delta H = \mu_0 \Delta H_0 + \frac{\alpha}{|\gamma|} f \quad (2)$$

where $\mu_0 \Delta H_0$ represents the inhomogeneous broadening of the FMR absorption peak arising from film nonuniformity. As shown in Figure 3(c) and 3(d), the fitted Gilbert damping constants are (2.4 ± 0.7) × 10$^{-3}$ for annealed GGG(5)/Pt(1)/YIG(30) sample, (2.0 ± 0.2) × 10⁻³ for annealed YIG(30) sample, and (5.4 ± 0.4) × 10⁻² for annealed NiO(8)/YIG(30) sample. The YIG film epitaxially grown on the GGG/Pt-nanoparticles spacer exhibits a magnetic damping reduced by a factor of approximately 22.5 compared with that of YIG grown on NiO, while maintaining a damping level comparable to that of YIG epitaxially grown on GGG substrate directly. This result confirms that the GGG/Pt-nanoparticles spacer effectively preserves the good crystalline quality and low damping characteristics of epitaxial T-YIG films.

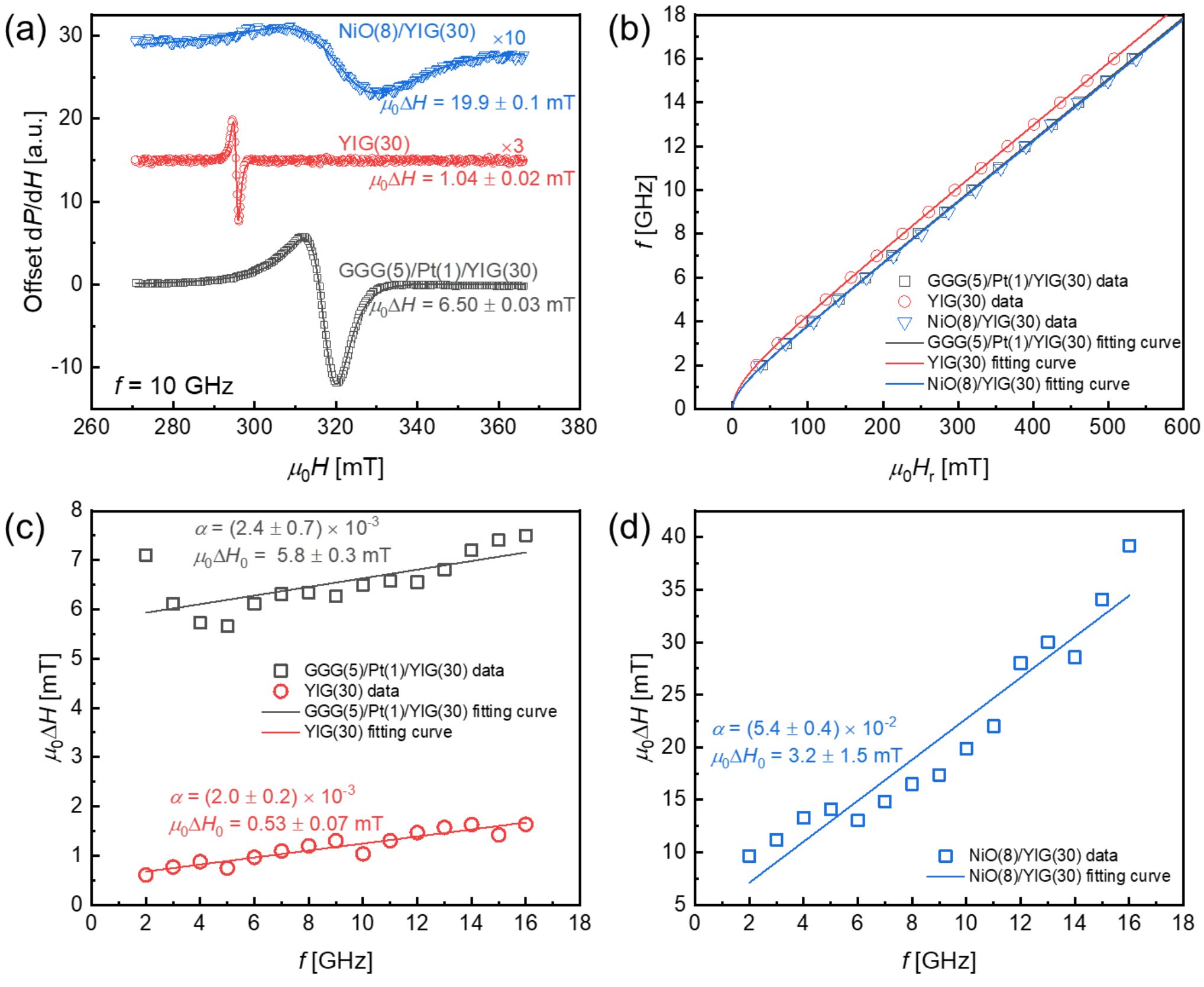

**Figure 3**. (a) FMR spectra of annealed GGG(5)/Pt(1)/YIG(30), YIG(30), and NiO(8)/YIG(30) films grown on GGG substrates measured at a microwave frequency of 10 GHz. The experimental data were fitted using the derivative of an asymmetric Lorentzian function to extract the linewidths $\mu_0\Delta H$. (b) The microwave frequency $f$ as a function of the resonance field $\mu_0 H_r$ for the three samples, fitted using the Kittel formula to determine the absolute gyromagnetic ratio $|\gamma|$. (c, d) Linear fitting of the linewidth $\mu_0\Delta H$ versus frequency $f$ for annealed GGG(5)/Pt(1)/YIG(30), YIG(30), and NiO(8)/YIG(30) samples, yielding the Gilbert damping coefficient $\alpha$ and the inhomogeneous broadening $\mu_0\Delta H_0$. The YIG film grown on the GGG/Pt-nanoparticles spacer exhibits a damping constant comparable to that of YIG directly epitaxially grown on GGG substrate, while showing a 22.5-fold reduction relative to the YIG film grown on NiO.

In the spin Seebeck effect (SSE) measurements, a distinct magnon junction effect in the YIG(40)/GGG($t_{\mathrm{GGG}}$)/Pt(1)/YIG(30) magnon junctions was observed. To detect the magnon current, a 3 nm Pt layer was deposited on top of the annealed YIG(40)/GGG($t_{\mathrm{GGG}}$)/Pt(1)/YIG(30) films using magnetron sputtering to serve as a spin detector via inverse spin Hall effect.[8-10] The device illustrated in **Figure 4**(a) was then fabricated through standard photolithography, etching, $SiO_2$ deposition, and lift-off processes. The top heating strip consists of a 3 nm Pt layer, while the connecting electrodes are composed of 10 nm Pt and 100 nm Au. The magnon junction effect is also demonstrated in Figure 4(a). When an electric current is applied to the Pt heater, a vertical temperature gradient across the film stack is generated. Due to the SSE, this temperature gradient induces a vertical magnon current through the magnon junction. By switching between the P and AP magnetic configurations of the two YIG layers, the transmitted magnon current can be effectively modulated, resulting in a corresponding change in the inverse spin Hall voltage detected in the top Pt strip.[8–10] As shown in Figure 4(b), the upper panel presents the magnetic hysteresis loop (black solid line) and its differential curve (red solid line), while the lower panel displays the SSE signal (black solid line) along with its differential curve (red solid line). Comparison between these two sets of data reveals that the magnitude of the inverse spin Hall voltage decreases when the magnon junction switch from the P state to the AP state. The differential SSE signal exhibits sharp peaks near zero magnetic field when switching from the P to AP state, indicating the observation of the magnon junction effect in the sample clearly. The critical switching fields of the T-YIG and B-YIG layers were identified from the peak positions in the differential curves, allowing the AP region to be highlighted (yellow background in Figure 4(b)). The device exhibits slightly higher switching fields compared to those obtained from VSM measurements on the raw film, which can be attributed to induced pinning centers of the YIG layers during the fabrication process, leading to a minor increase in coercivity.

To further verify that the measured signals originate from SSE, we performed heating-current-dependent and magnetic-field-angle-dependent measurements. As shown in Figure 4(c), we measured the SSE voltage response of the GGG//YIG(40)/GGG(5)/Pt(1)/YIG(30) sample in the positively saturated magnetic state (P state) under external magnetic fields, while varying the heating current from –9 mA to +9 mA. The resulting SSE voltage exhibits a clear quadratic dependence on the applied current, which can be well fitted by a parabolic curve of the heating current, consistent with the characteristic behavior of SSE signals.[5][8] Figure 4(d) presents the angular dependence of the SSE voltage measured under a 1 T magnetic field applied to the sample. The current direction is defined as the +X axis, and the rotation angles $\theta$, $\zeta$, $\phi$ correspond to field rotations in the XOZ, YOZ, XOY planes, respectively, as illustrated in the upper schematic of Figure 4(d). The SSE voltage is nearly independent of $\theta$, but exhibits $\cos\zeta$ and $-\sin\phi$ dependences for the $\zeta$ and $\phi$ rotations, respectively, consistent with the characteristic angular symmetry of the SSE signal.[5][8] Figure 4(e) summarizes the SSE measurement results for samples with $t_{\mathrm{GGG}}$ = 5, 6, and 7 nm. In all cases, the differential curves exhibit four peaks corresponding to the switching between P and AP magnetic configurations, further confirming the reproducibility of the magnon junction effect across different GGG spacer thicknesses.

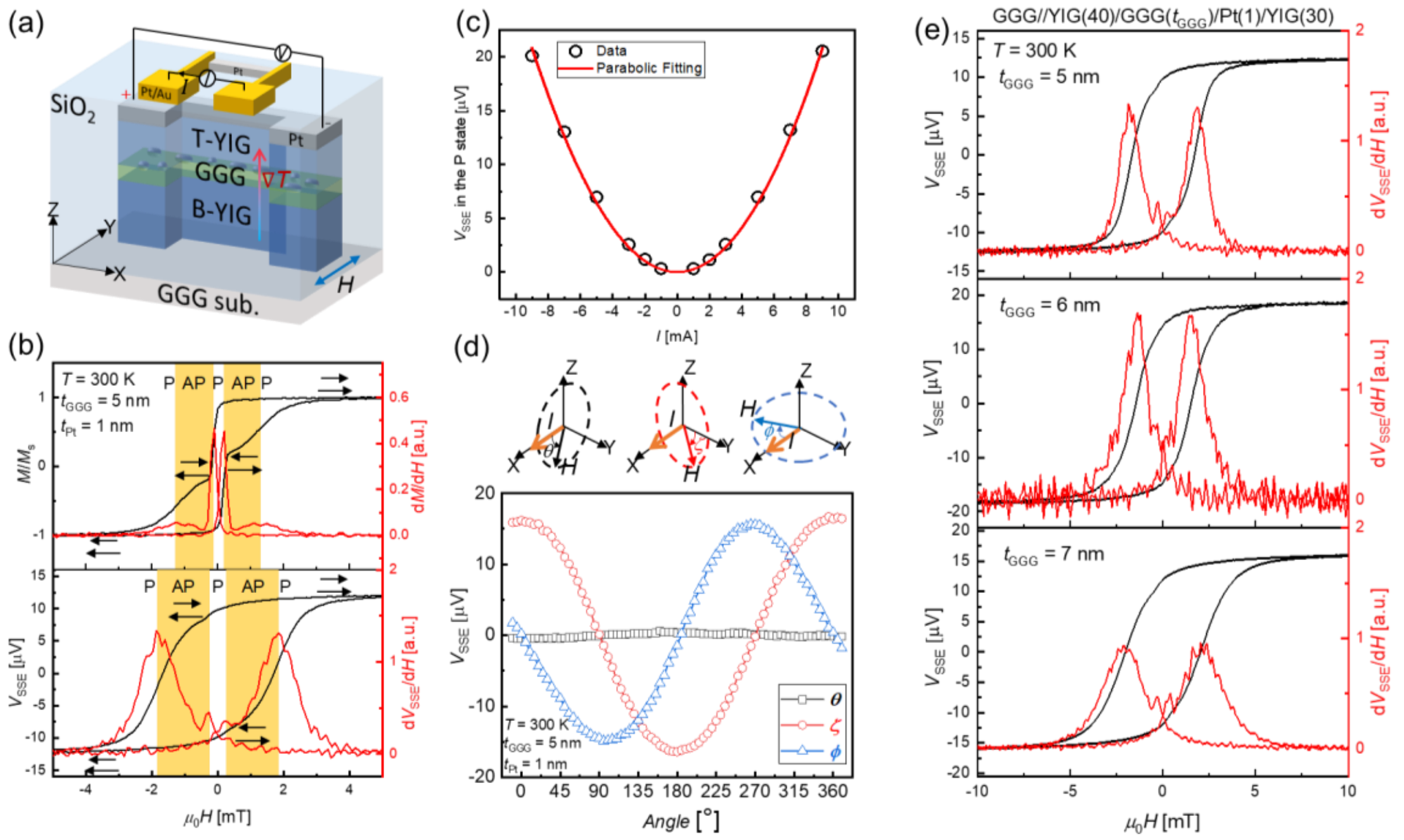

**Figure 4**. (a) Schematic illustration of the SSE measurement device and configuration. (b) Comparison between the magnetic hysteresis loop (upper panel) and the SSE measurement result (lower panel) for the GGG//YIG(40)/GGG(5)/Pt(1)/YIG(30) device at 300 K. (c) SSE voltage of the GGG//YIG(40)/GGG(5)/Pt(1)/YIG(30) device under heating currents ranging from −9 to 9 mA at 300 K. The voltage signal can be well fitted by a parabolic function of the heating current, consistent with the SSE characteristics. (d) Angular dependence of the SSE voltage measured at 300 K under an external magnetic field of 1 T for the same device. The measurement geometry for field rotations in the XOZ, YOZ, XOY planes is illustrated in the upper schematic. (e) SSE signals at 300 K for devices with $t_{\mathrm{GGG}}$ = 5, 6, 7 nm and fixed $t_{\mathrm{Pt}}$ = 1 nm. All exhibit signatures of the magnon junction effect.

We further investigated the temperature dependence of the SSE signal in samples with different GGG spacer thicknesses. **Figure 5**(a)-(c) present the SSE voltage data at different temperatures for $t_{\mathrm{GGG}}$ = 5, 6, and 7 nm, respectively, while Figure 5(d)-(f) show their corresponding differential curves. The coercivity of YIG films is expected to increase monotonically as temperature decreases.[31] This trend was also confirmed by temperature-variable VSM measurements performed on separately prepared and annealed YIG(40) and GGG(5)/Pt(1)/YIG(30) samples grown on (111)-oriented GGG substrates, as shown in **Figure S4**. As indicated by the dashed lines in Figure 5(f), for $t_{GGG}$ = 7 nm, the coercive fields of T-YIG layers follow this expected monotonic increase with decreasing temperature. However, when $t_{GGG}$ = 5 and 6 nm, the coercivity of the T-YIG layer initially increases, then decreases, and finally increases again upon cooling, as indicated by the dashed lines in Figures 5(d) and 5(e). This anomalous behavior suggests that the coupling between the two YIG layers becomes stronger as temperature decreases. At room temperature, the coupling is weak, and the coercivity of the T-YIG layer increases as the temperature decreases. With further cooling, the coupling between two YIG layers strengthens. Because the B-YIG layer has a smaller coercive field and switches first, it then exerts an effective coupling field on the T-YIG, causing it to reverse at a smaller applied field — thus leading to an apparent reduction in coercivity of the T-YIG. Upon further temperature decrease, the two YIG layers become more strongly coupled and switch together, resulting in a renewed increase in the overall coercivity. As is evident in Figures 5(d) and 5(e), within the temperature range in which the coercivity of the T-YIG decreases with cooling, the AP magnetic configuration gradually disappears, indicating the emergence of strong coupling. In contrast, when $t_{GGG}$ reaches 7 nm, as shown in Figures 5(f), the coupling between the two YIG layers remains weak throughout the 50 K - 300 K temperature range, allowing the AP state to continuously persist.

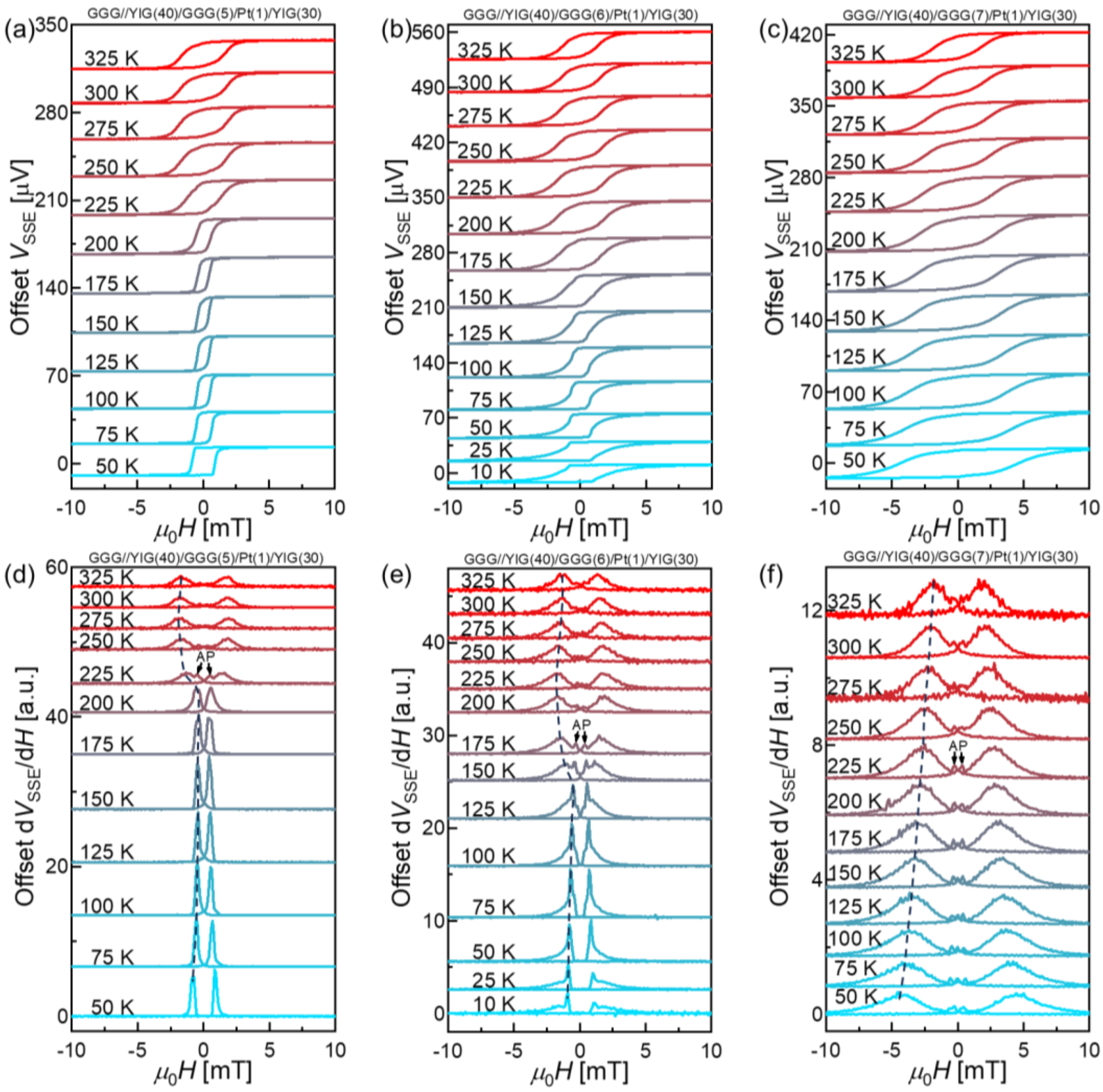

**Figure 5**. Temperature-dependent SSE measurements for samples with fixed $t_{\mathrm{Pt}}$ = 1 nm and $t_{\mathrm{GGG}}$ = (a, d) 5 nm, (b, e) 6 nm, and (c, f) 7 nm. (a–c) show the SSE voltage data, while (d–f) present the corresponding differential curves. As an illustration, the peaks indicating the appearance of the AP state in several differential curves are marked with arrows. The temperature-dependent evolution of the switching field of the T-YIG layer is also indicated by dashed lines.

To quantitatively evaluate the strength of the magnon junction effect, the magnon junction ratio (MJR) is defined as $\eta_{\mathrm{mj}} = (V_{\mathrm{P}} - V_{\mathrm{AP}})/(V_{\mathrm{P}} + V_{\mathrm{AP}})$.[16] Compared with the definition $V_{\mathrm{P}}/V_{\mathrm{AP}}$, this definition is non-divergent, with values closer to 100% indicating a more complete switching-off of magnon transport, which facilitates performance comparison. $V_{\mathrm{AP}}$ represents the SSE voltage measured at the magnetic field corresponding to the differential minimum between the two differential peaks which indicate the appearance and disappearance of the AP state, and $V_{\mathrm{P}}$ denotes the SSE voltage at the saturation field under the P state. Among all fabricated samples, the MJR reaches a maximum value of 43% at 225 K on the YIG(40)/GGG(5)/Pt(1)/YIG(30) sample, as shown in **Figure 6**(a), where the orange squares on the black curve indicate the position where the $V_{\mathrm{AP}}$ was obtained. The extracted MJRs for different GGG thicknesses and temperatures are summarized in Figure 6(b) (solid symbols and lines). When the AP state vanishes due to strong interlayer coupling, the MJR becomes undefined and is therefore omitted from the plot. Interestingly, for samples with $t_{\mathrm{GGG}}$ = 5 and 6 nm, the MJR shows a nonmonotonic evolution with decreasing temperature, first decreasing and then increasing, which suggests a possible correlation with the coupling between the two YIG layers. Based on the differential SSE curves shown in Figure 5(d–f), the temperature-dependent coercivities of the T-YIG layers were extracted and are presented in Figure 6(b) (open symbols and dashed lines). A direct comparison reveals that the increase in MJR coincides with the coupling-induced decrease in the coercivity of the T-YIG layer upon cooling, with both effects occurring in the $t_{\mathrm{GGG}}$ = 5 and 6 nm samples within the same temperature range. In contrast, when $t_{\mathrm{GGG}}$ reaches 7 nm, the coupling between the two YIG layers remains weak always, and the MJR decreases monotonically with decreasing temperature. This correlation indicates that the interlayer magnetic coupling can enhance the magnon junction effect.

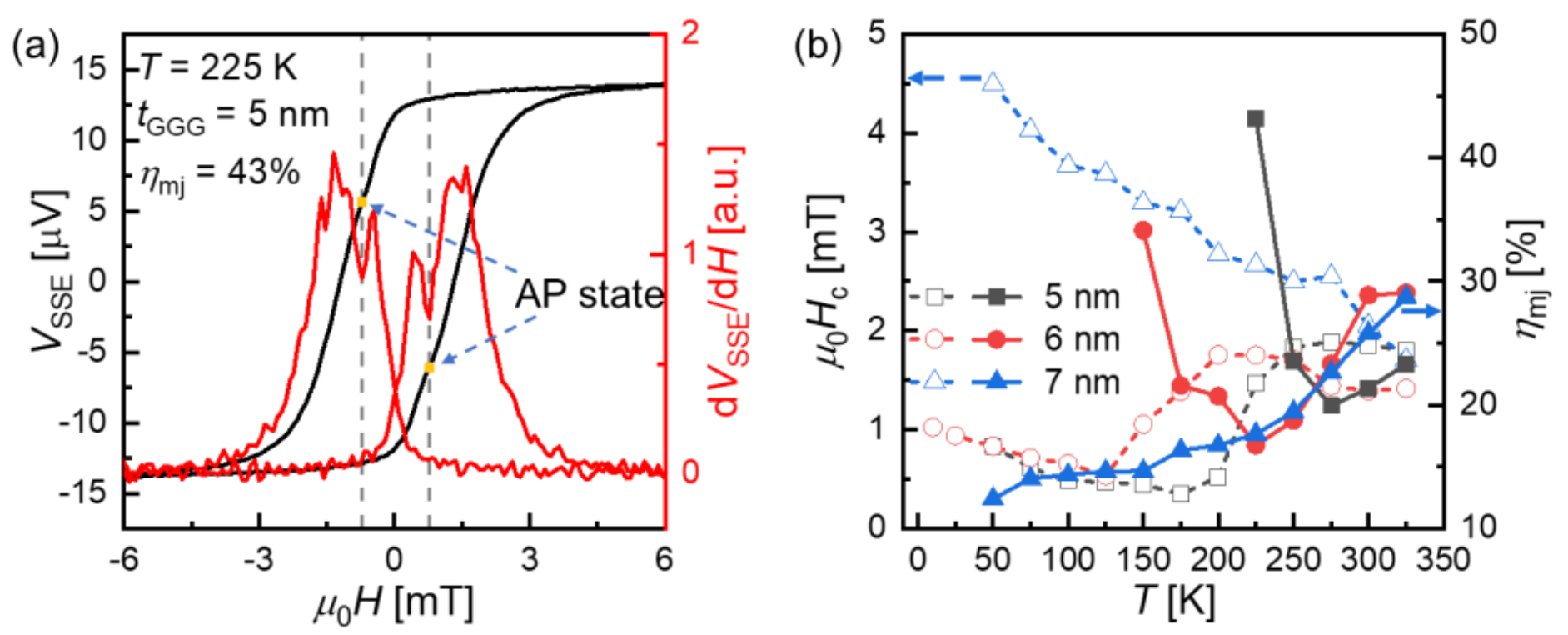

**Figure 6**. (a) SSE measurement of the YIG(40)/GGG(5)/Pt(1)/YIG(30) magnon junction showing an MJR $\eta_{\mathrm{mj}}$ of 43% at 225 K. The orange squares mark the AP state where the $V_{\mathrm{AP}}$ is determined. (b) Temperature dependence of the $\eta_{\mathrm{mj}}$ (solid symbols and lines) and the coercive field of the T-YIG layer $\mu_0 H_c$ (open symbols and dashed lines) for samples with $t_{\mathrm{GGG}}$ = 5, 6, 7 nm and fixed $t_{\mathrm{Pt}}$ = 1 nm.

## 3. Conclusion

In summary, a fully epitaxial magnon junction have been fabricated and optimized employing a hybrid GGG/Pt-nanoparticles spacer, and an MJR of 43% was observed. Compared with previously reported magnon junctions employing non-epitaxial spacers such as NiO, this structure preserves superior crystalline quality, thereby reducing magnon current loss. Temperature-dependent measurements further reveal that interlayer coupling can enhance the MJR. This work provides valuable insights toward the realization of high-performance magnonic devices with enhanced functionality.

## 4. Methods

*Material Growth*: The YIG(40)/GGG($t_{\mathrm{GGG}}$)/Pt($t_{\mathrm{Pt}}$)/YIG(30) heterostructures were grown on (111)-oriented GGG substrates with dimensions of 10 mm × 10 mm × 0.5 mm using magnetron sputtering at room temperature. The YIG and GGG layers were deposited by RF sputtering at an Ar pressure of 1.5 Pa, with RF powers of 80 W and 100 W, respectively. The Pt layer was deposited by DC sputtering at an Ar pressure of 0.16 Pa with a power of 100 W. After deposition, the films were annealed in air at 800 °C for 80 minutes in a muffle furnace, with heating and cooling rates of 9.75 and −9.75 °C $\mathrm{min}^{-1}$, respectively. The NiO layer in the NiO(8)/YIG(30) control sample was deposited by RF sputtering at an Ar pressure of 0.55 Pa with a power of 80 W.

*HADDF imaging and EELS measurements*: High-quality thin-cross-sectional samples were prepared using the focused ion beam (FlB, Helios 600i from FEl). The HAADF images and the EELS spectrum were collected on a JEOL ARM200F NEOARM equipped with a double spherical aberration corrector at 200 kV. The HAADF images were collected at acceptance angles of 90-370 mrad. The EELS spectrum was collected in the line scan mode at a collection angle of ≈23 mrad and a convergence angle of ≈25 mrad.

*VSM measurements*: The magnetic hysteresis loops shown in Figure 2 and S3 were measured using a calibrated VSM (EZ-9 from MicroSense). The low-temperature magnetic hysteresis loops in Figure S4 were measured by VSM option of a Physical Property Measurement System (PPMS, Quantum Design PPMS-9H),

*FMR measurements*: During FMR measurements, each film was placed face-down on a coplanar waveguide. Microwaves with frequencies ranging from 2 to 16 GHz were applied through the waveguide by a microwave signal generator (SMB 100A from Rohde&Schwarz), and the transmitted signal amplitude was detected by a Schottky diode detector at the end of the circuit and fed into a lock-in amplifier (SR865A from Stanford Research Systems). An in-plane external magnetic field perpendicular to the microwave magnetic field was applied, consisting of a DC component supplied by an electromagnet and a 133.77 Hz low-frequency AC component generated by an additional Helmholtz coil. The low-frequency AC current applied to the Helmholtz coil was supplied by an AC current source (Keithley 6221), which also provided the reference signal for the lock-in amplifier. The amplitude of the AC current and the microwave power were carefully adjusted to avoid spectral distortion and to ensure a high signal-to-noise ratio for different samples.

*Magneto-thermoelectric transport measurements*: The SSE measurements were performed in a PPMS, with the heating current supplied by a source meter (Keithley 2400) and the inverse spin Hall voltage detected using a nanovoltmeter (Keithley 2182).

This manuscript was linguistically polished using ChatGPT (OpenAI), under the full supervision of the authors. The use of the tool was limited to language editing, and no scientific interpretation or data analysis was generated by the AI system.

**Acknowledgements**

This work was financial supported by the National Key Research and Development Program of China [MOST Grant No. 2022YFA1402800], the National Natural Science Foundation of China [NSFC, Grants No. T2495210, 12134017, 12374131, and W2412079], and partially supported by the Chinese Academy of Sciences President's International Fellowship Initiative (PIFI Grant No. 2025PG0006).

**Supporting Information for**

# Epitaxial Magnon Junctions with $Y_3Fe_5O_{12}$/$Gd_3Ga_5O_{12}$/Pt-nanoparticles/$Y_3Fe_5O_{12}$

*Jianing Lin, Haoyu Zhuang, Tianyi Zhang, Jiyang Ou, Jihao Xia, Hongyu Wang, Jiang Niu, Qinglong Yu, Caihua Wan*, Guoqiang Yu, Richeng Yu, and Xiufeng Han**

E-Mail: xfhan@iphy.ac.cn; wancaihua@iphy.ac.cn

1. **Electrical transport measurement on the annealed GGG//YIG(40)/GGG(5)/Pt(1)/YIG(30) sample.**

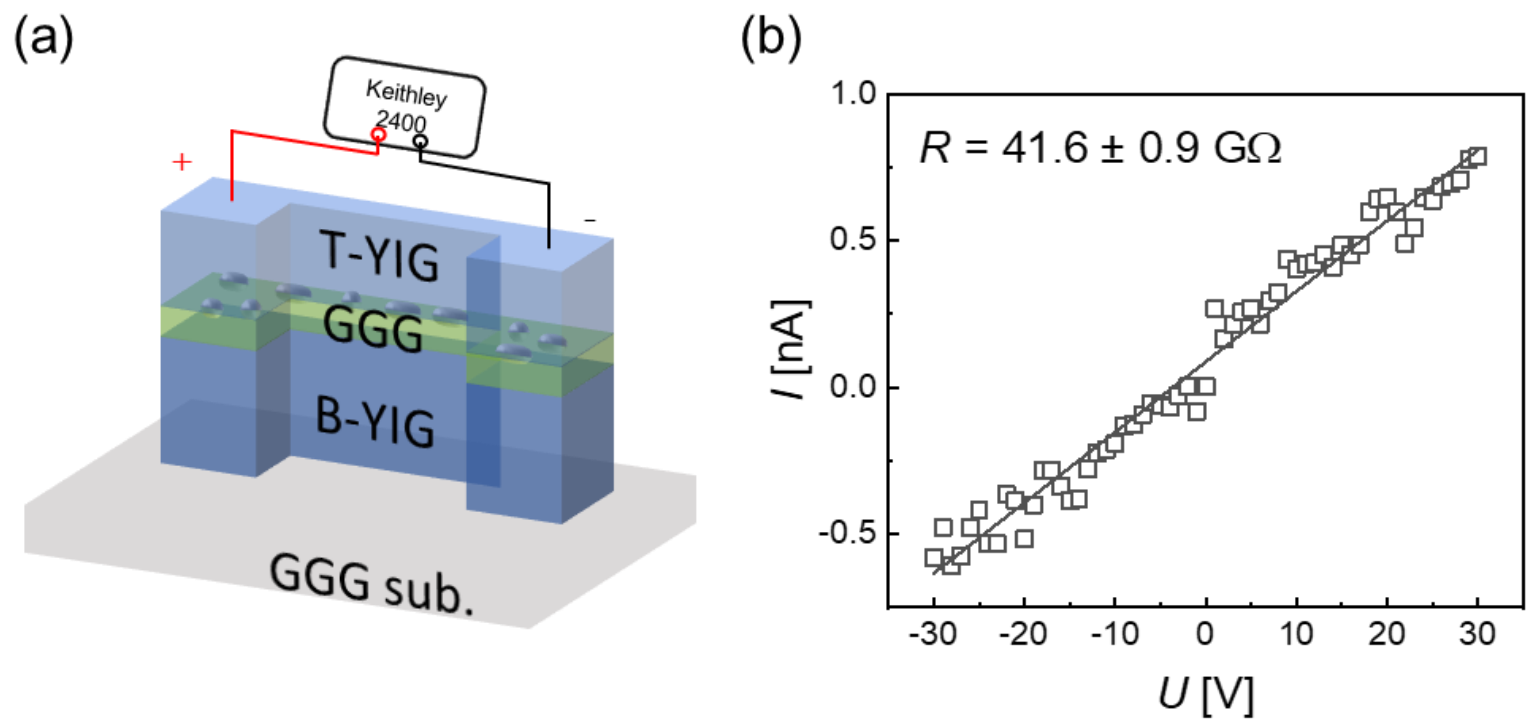

**Figure S1**. (a) Schematic illustration of the electrical transport measurement setup for the annealed GGG//YIG(40)/GGG(5)/Pt(1)/YIG(30) sample. Static voltages are applied using a Keithley 2400 source meter, which simultaneously measure the resulting currents. (b) Measured current as a function of the applied voltage, yielding a resistance of 41.6 ± 0.9 GΩ, confirming the highly insulating nature of the annealed film.

2. **Local HAADF image and EELS elemental mapping of the annealed GGG//YIG(40)/GGG(5)/Pt(1)/YIG(30) film.**

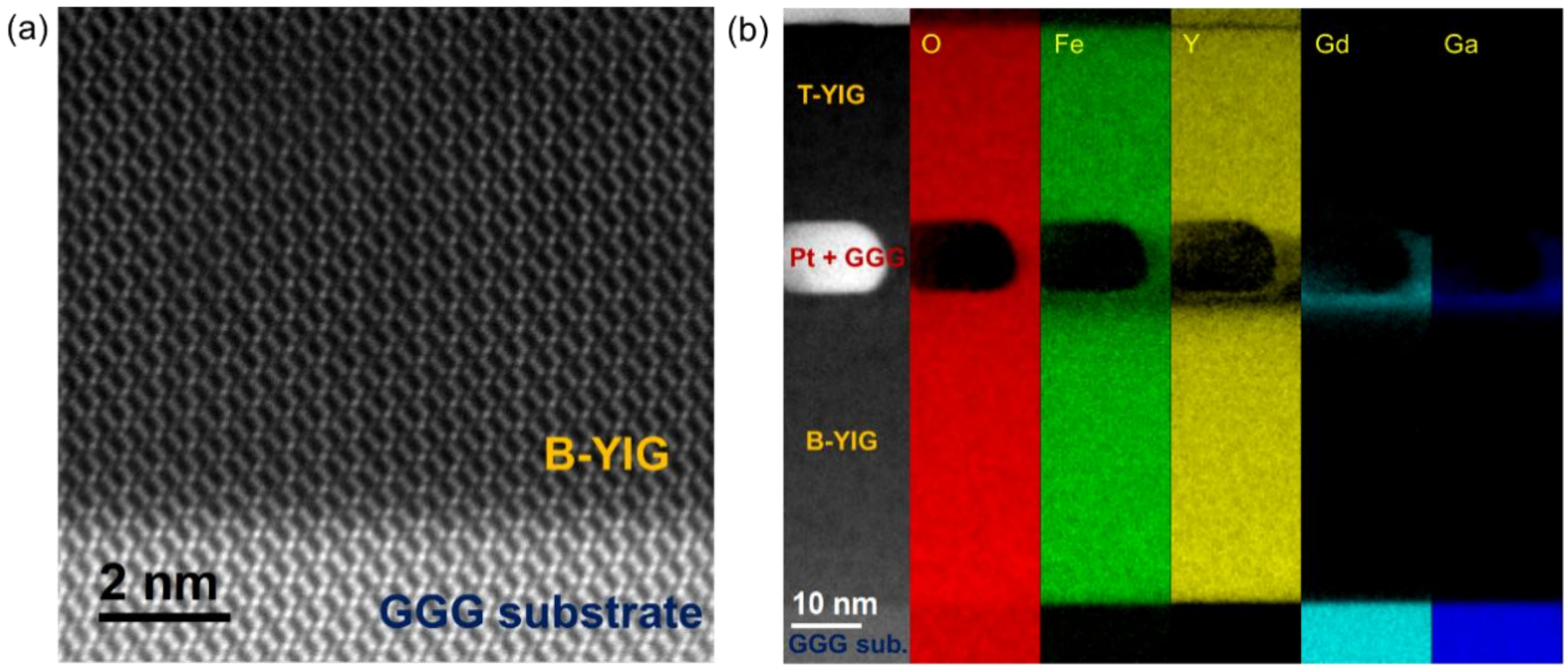

**Figure S2**. (a) HAADF image of the annealed GGG//YIG(40)/GGG(5)/Pt(1)/YIG(30) film, highlighting the interface between the B-YIG layer and the GGG substrate. (b) EELS elemental mapping of the sample, showing a sharp interface between the GGG

substrate and the B-YIG layer, as well as noticeable elemental interdiffusion between the GGG spacer and the T-YIG layer.

## 3. VSM measurements of the annealed GGG//YIG(40)/GGG($t_{GGG}$)/YIG(30) films.

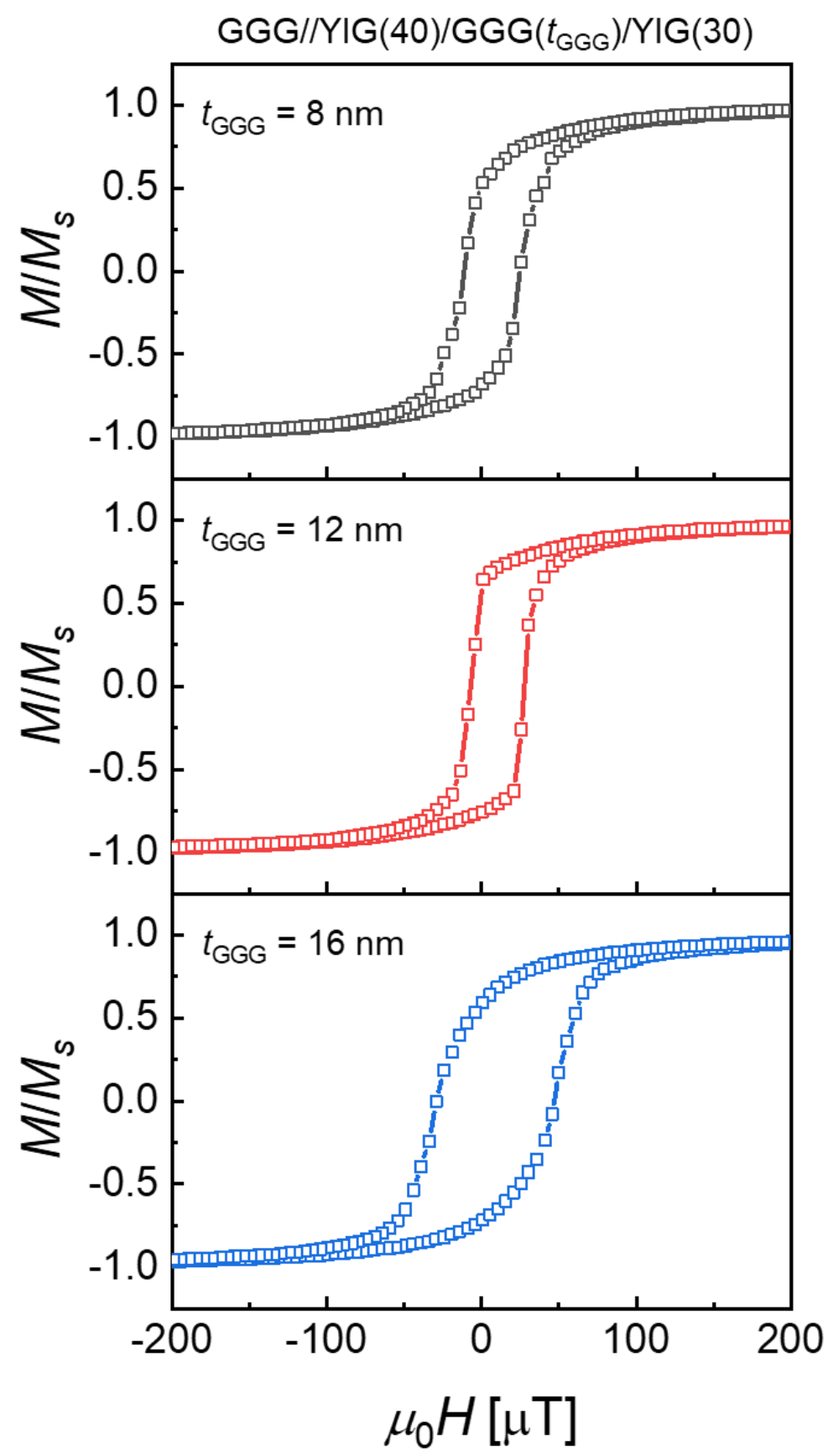

**Figure S3**. VSM measurement data for GGG//YIG(40)/GGG($t_{GGG}$)/YIG(30) films with $t_{GGG}$ = 8, 12, and 16 nm. Both T-YIG and B-YIG layers exhibit very small coercive fields, preventing the establishment of a well-defined AP state.

## 4. Temperature-dependent VSM measurements of annealed GGG//YIG(40) and GGG//GGG(5)/Pt(1)/YIG(30).

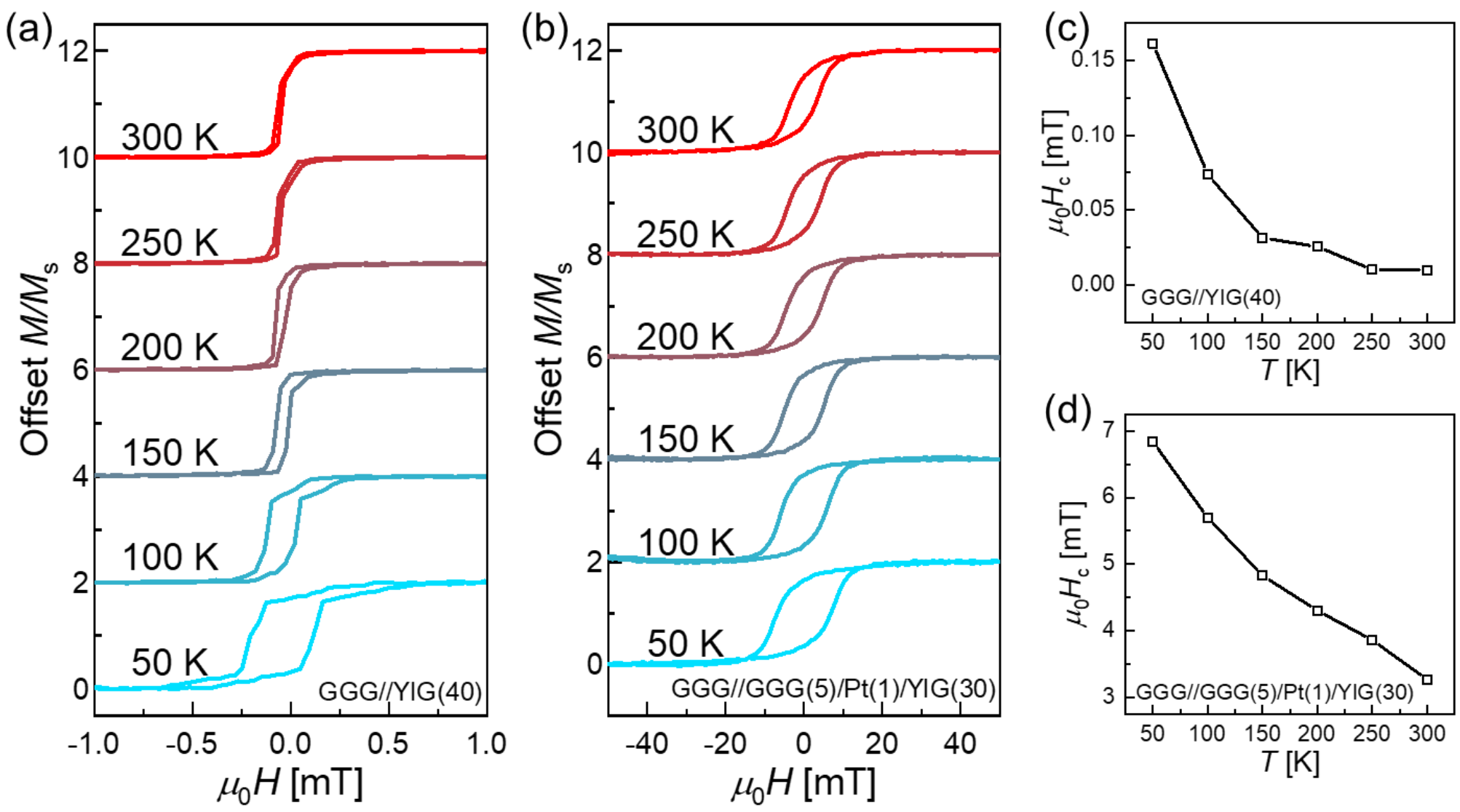

**Figure S4**. Temperature-dependent VSM measurement data for (a) GGG//YIG(40) and (b) GGG//GGG(5)/Pt(1)/YIG(30) samples. The variations of coercivity with temperature for the two samples are summarized in (c) and (d), respectively. In both cases, the coercive field increases monotonically as the temperature decreases.